\documentclass[aps,twocolumn,amssymb,showpacs]{revtex4}
\usepackage{graphicx}

\begin{document}

\def\be{\begin{equation}}
\def\ee#1{\label{#1}\end{equation}}
\title{ Transition from accelerated to decelerated regimes in JT and CGHS
cosmologies}
\author{M. B. Christmann, F. P. Devecchi\footnote{devecchi@fisica.ufpr.br},
G. M.  Kremer\footnote{kremer@fisica.ufpr.br}, and C. M. Zanetti}
\affiliation{Departamento de F\'\i sica, Universidade Federal do Paran\'a,
Caixa Postal 19044, 81531-990, Curitiba, Brazil}

\begin{abstract}

In this work we discuss the possibility of positive-acceleration regimes,
and their transition to decelerated regimes, in
two-dimensional (2D) cosmological models.
We use general relativity and the thermodynamics
in a 2D space-time, 
where the gas is seen as the sources of the gravitational
field.
An early-Universe model is analyzed where the  state equation of van der Waals
is used,
replacing the usual barotropic equation. We show that this substitution
permits 
 the simulation of  a period of inflation, followed by a 
negative-acceleration era. The dynamical behavior
 of the system
follows from  the solution of the Jackiw-Teitelboim equations (JT equations)
and the energy-momentum conservation laws.
In a second stage we focus the  Callan-Giddings-Harvey-Strominger model (CGHS
model); here the transition from the
inflationary period to the decelerated period is also present between
 the solutions, although
this result depend strongly on the initial conditions used for the dilaton
field. The temporal evolution of the cosmic scale
function, its acceleration, the energy density and the 
hydrostatic pressure are the physical quantities obtained in through 
the analysis.

\end{abstract}
\maketitle

\section {Introduction} 

The proposal of theories of gravity in lower dimensions has 
been 
studied intensively~\cite{Bro}. 
These models offer interesting results that, 
if properly
analyzed, can be used  to solve problems in realistic theories. 
 Cosmological solutions  
has been also under analysis in these formulations~\cite{RMann,Dese,Cado}.
Taking the particular case of two-dimensional (2D) theories,  the gravity
 model proposed by   
  Teitelboim  and Jackiw~\cite{Jackiw,Teit} (JT model) provided 
consistent results at the  classical and quantum levels~\cite{Bro}. 
This model   works as the ``closest counterpart''  of general
relativity in 2D formulations, since the Einstein field 
equations furnish no dynamics in 2D~\cite{Bro}. In the JT model a 
scalar field (related to the dilaton~\cite{Jackiw}) is present. In the 
``Einsteinian'' interpretation the dilaton is
working solely as an auxiliary variable to guarantee the correct equations of
motion for the geometrical quantities~\cite{Bro}. Another
analysis of the JT model, in a cosmological context~\cite{Cado}, considered
the dilaton as a dynamical field. 
The best cosmological results obtained using the JT model include 
the description of a Universe 
filled with ordinary  matter or/and electromagnetic
radiation, where its behavior is described in terms of the temporal 
evolution of the scale parameter, the energy density and the total
pressure~\cite{RMann,Cado,K-D}. A remarkable point here is that positive
 accelerated regimes (such as an inflationary Universe)
were possible only under 
very special conditions, such as a negative vacuum energy density for the
inflaton and violation of the weak energy condition for the matter/radiation
constituents~\cite{RMann}.

Inspired in string theory, Callan et al.~\cite{CGHS} proposed the
 so-called CGHS model. 
In this theory the dilaton is invariably a
dynamical field; in fact this model is formally related to the JT model,
through a field mapping, when we consider the dilaton  as part of the
true degrees of freedom of the gravitational field~\cite{Jackiw}.
 Here the cosmological results include, again, the 
description of a matter/radiation filled Universe~\cite{RMann}. The
possibility of description of  inflation or
dark energy regimes depend  basically on  the same special 
conditions present in
the JT model~\cite{RMann}.

In this work, as a sequence to our previous effort~\cite{K-D},
 we want  to
 discuss the existence of positive-acceleration 
solutions in 2D cosmologies; these related  to the description of an
 inflation period and its
transition to a decelerated  period.
 We show that a  consistent formulation is possible 
when the usual barotropic  
equation of state is replaced  by the van der
Waals equation (VDW), as was proposed by Capozziello et al.~\cite{Capo} 
in 4D models and explored by one of these authors~\cite{Kre1}.
We show that  an inflationary scenario is possible in the JT model, with a
compatible behavior of the physical quantities. In fact, we obtain
the temporal evolution of the scale factor, its acceleration, and the energy
density of the VDW constituent.  We also
obtain an accelerated regime, with a transition to a decelerated era, 
 in the CGHS model,
although in this case the behavior of the acceleration field depends
 strongly on the  initial conditions of the dilaton field.

The manuscript is structured as follows. In Section II we make a brief
review of JT and CGHS models in a cosmological context. In Section III, 
accelerated regimes
with their transition to deceleration are focused in the JT
model when we take into account the VDW equation of state.
Moreover, we describe positive
accelerated regimes in the CGHS model,  using again the VDW equation,
together with a discussion of the results.  Units have been chosen so that 
$G=c=k=1$.

\section{Cosmology in  JT and  CGHS models}

In this section we make a brief review of the JT and CGHS models
in a cosmological context, focusing on the results present 
in~\cite{RMann,Cado,K-D}.

One fundamental point in 2D gravity is that
the relation between the metric tensor and the sources 
is modified. This is because, as it is well-known, the Einstein field 
equations give no     
dynamics in the 2D case~\cite{Bro}; in fact this  is a consequence 
of general coordinate transformation and conformal 
invariance~\cite{Bro}.

Another important feature of 2D models
is that they show
considerable less mathematical complexity and at the same time they
preserve
the gauge principles that are used to construct their 4D
counterparts. One impressive result is that in 2D models the 
quantization of the 
gravitational field is possible~\cite{Bro}, opening
the possibility, for the first time, between other perspectives, 
of full quantum
cosmological models for the very early Universe.~\cite{Bro,RMann}.

The model proposed by Teitelboim and Jackiw ~\cite{Jackiw,Teit} 
focused on the fact that in 2D the full geometrical 
information for the 2D space-time is encapsulated in the curvature scalar $R$. 
The  action for this
model is given by
\begin{equation}
S= \int d^2x \sqrt{-g}\left\{N(x)\left[R(x)-\Lambda+8\pi
T^\mu_\mu(x)\right]\right\},
\end{equation}
where $T^\mu_\mu(x)$ is the trace of the energy-momentum
tensor of the sources and $\Lambda$ is a cosmological constant. 
Using the variational
principle for the scalar  field  $N(x)$ (which is related to the dilaton
~\cite{Jackiw})  the equation of motion that
follow is
\be
R (x) = -8\pi T^\mu_\mu(x)+\Lambda.
\ee{RW21a}

When we apply the variational principle to the components of the metric
we obtain a  relation that {\it defines} the scalar field $N(x)$ in terms 
of the other fields. In 
this strict  ``Einsteinian'' interpretation the scalar field is taken as an 
auxiliary
field; that means, this formulation is the closest general relativity 
counterpart in 2D. On the other hand,
 in a ``Brans-Dicke'' interpretation, the scalar field $N(x)$ is
dynamical~\cite{Cado}. We will consider the dilaton as a 
dynamical field  in the CGHS model
(see end of this and next section).

In this hypothetical 2D Universe the assumptions of spatial homogeneity 
and isotropy are also invoked. The
Robertson-Walker metric has  the
following form in a 2D Riemannian space 
\be 
ds^2=(dt)^2-a(t)^2(dx)^2,
\ee{RW1a}
where  
$a(t)$  is the  cosmic scale factor, that in this case encloses 
the complete information
about the  evolution of the gravitational field created by the sources. In
fact the usual geometrical quantities (Ricci tensor, curvature scalar) turn
out to be
\be
R_{00}={\ddot a\over a},\qquad
R_{11}=-\ddot a a,
\qquad
R=2{\ddot a\over a}.
\ee{RW21}

The cosmological solutions of the JT model  include a Universe filled with 
radiation or
matter~\cite{RMann,K-D}; the equations of motion (together with the
conservation law ${T^{\mu\nu}}_{;\nu} =0$ for a perfect fluid 
where $T^\mu_\nu=\hbox{diag}(\rho,-p)$)  can be expressed as
\be
{\ddot a\over a}=-4\pi(\rho-p)+{\Lambda\over2},
\ee{98}
\be
\dot \rho+{\dot a\over a}(\rho+p).
\ee{99}
The system of differential equations above can be solved 
for $a(t)$ and $\rho(t)$ for given initial conditions, once the 
equation of state $p=p(\rho)$ is specified.

Inflationary scenarios appear only when unusual hypothesis are taken into
account.  Negative energy density and violation of the weak energy
condition~\cite{RMann} are the most representative;
 on the other hand, new features appear in 
the cosmological 
solutions when
the auxiliary field is taken as a dynamical variable~\cite{Cado}. In fact,
this approach is formally linked to the analysis of the so-called CGHS
model~\cite{CGHS}. 

The CGHS model  was proposed initially for the physical investigation of
2D black-holes ~\cite{CGHS}.
The action, inspired in string theories~\cite{CGHS},
 includes  the dilaton field 
in a similar way to the 4D Jordan-Brans-Dicke theories (although a complete 
analogy cannot
be done~\cite{Cado}). As was mentioned before, the dilaton is seen as a
 dynamical field in this case. The equations of motion are
\be
e^{-2\phi}\left[R_{\mu \nu} -\Lambda g_{\mu\nu}- 2 \nabla _{\mu} 
\nabla _{\nu} \phi \right] =
-8\pi T_{\mu \nu},
\ee{100}
\be  
R -\Lambda- 4(\nabla \phi )^2 + 4 \nabla ^2 \phi  = 0.
\ee{101}   
where $\phi $ is the dilaton and $T_{\mu \nu}$ is the
energy-momentum tensor of the sources.
For a perfect fluid constituent these gravitational field equations read
\be
{{\ddot a} \over a} = -4\pi e^{2\phi } ( \rho -
p)+ {{\dot a \dot \phi }\over a} +\ddot \phi+{\Lambda\over 2},
\ee{102}
\be
{{\ddot \phi} } = 4\pi e^{2\phi} (\rho  + p)
+ {{\dot a \dot \phi }\over a}.
\ee{103}
If we know the equation of state $p=p(\rho)$ and prescribe initial conditions,
it is possible to obtain
from the system of differential equations (\ref{99}), (\ref{102}) and 
(\ref{103}) the evolution equations for the fields $\rho(t)$, $a(t)$ 
and $\phi(t)$.

The  regimes coming out include dust and radiation filled Universes with the
possible inclusion of a cosmological constant
~\cite{RMann,Cado}. The existence of positive accelerated
solutions depend, as in the JT model, on the imposition of  unusual
conditions as negative energy density for the scalar field that represents
the inflaton.

In the following section we present a solution to this problem considering
the van der Waals state equation (VDW)~\cite{Capo} to model the
 behavior of an early Universe.

\section{Inflation in JT and CGHS models}

In this section we consider the Jackiw-Teitelboim model (JT model), 
to describe 
an inflationary scenario in 
a 2D Universe. The scalar
field  $N(x)$, mentioned in the precedent section, is taken as an 
auxiliary field. This Universe is seen as a fluid 
with its thermodynamical
state ruled by a van der Waals (VDW) equation. This state equation was
considered in a cosmological context in the works~\cite{Capo,Kre1}. 
 In classical thermodynamics, the VDW equation is a
refinement of the ideal gas equation of state, 
when the volume of particles and the long range interaction
between  the particles is
taken into account for a dense fluid. In a 2D cosmological context  
we shall neglect the  term related to the long range interaction 
and write the VDW equation of state as
\begin{equation}
p={w\rho\over 1-\alpha\rho},
\end{equation}
where $w$ and $\alpha$ are constants. When we introduce the above 
expression in equations (\ref{98}) and (\ref{99})
 we get
\be
{\ddot a\over a}=-4\pi\left(\rho-{w\rho\over 1-\alpha\rho}\right)
+{\Lambda\over2}, \quad
\dot\rho+{\dot a \over a}\left(\rho+{w\rho\over 1-\alpha\rho}\right)=0.
\ee{3}

The system of differential equations (\ref{3}) is solved numerically by 
prescribing the following initial
conditions:  $a(0)=1$, $\dot a(0)=1$ and $\rho(0)=1$. Moreover,
in order to plot the figure 1, 
we have chosen  $\alpha=0.5$, $\Lambda=0.002$ and
two values for $w$, namely, 0.9 and 0.8. The time evolution
of the physical quantities obtained show  (see figure 1) an
expansion, with a transition from a positive accelerated regime, that
would correspond to the inflationary period, followed by a 
decelerated period where the VDW equation approaches a barotropic 
equation of state, since the energy density is decreasing. At later 
times the Universe returns to an accelerated epoch, owing
to the presence of the cosmological constant. Whereas the behavior of 
the acceleration field
in 2D is analogous to the one obtained in four dimensions~\cite{Kre1}, 
the physical interpretations are quite different. In 4D the transition from an 
early accelerated regime to a decelerated epoch is due to the fact 
that the pressure of the VDW fluid  changes from a negative value, where 
it behaves like an inflaton, to a positive value, where it behaves like
a matter dominated fluid. In the 2D case, the pressure of the VDW fluid 
in the earliest times is positive and larger than the energy density 
whereas at later times it follows a barotropic equation of state. 
The transition from a decelerated regime 
to an accelerated epoch is due to a presence of a cosmological constant
which has also different interpretations; 
while in the 4D case the cosmological 
constant plays the role of a dark energy this does not occur in the 2D 
case, since  the sign of the energy density in the acceleration 
equation (\ref{98}) is negative, and a negative cosmological constant 
would contribute more for the deceleration of the  2D Universe. 
If one considers a vanishing cosmological constant in the 2D case 
the deceleration of the Universe  leads to a "Big Crunch". This last result 
was also obtained in the work~\cite{K-D} where a barotropic equation of 
state was used to model the cosmological fluid.
For a fixed value of the parameter $w$  that permit the transition from
an accelerated regime to a decelerated one (for $\alpha = 0.5$ the critical
value is $w=0.98$) the results show that the 
smaller the
value of $\alpha $ the more drastic is the transition. 
Furthermore, 
for a fixed $\alpha$ the smaller the value of $w$ the transition to the
decelerated regime is more pronounced (see figures 1 and 2).  
In all cases, after the transition
occurs, the accelerated regime returns only when a  cosmological
constant term 
is present. 
On the other hand, for big values of the time variable  the
scale factor shows a collapsing  Universe, dominated by a constituent with
a barotropic equation of state when the cosmological constant term is absent. 

\begin{figure}\vskip0.8truecm\begin{center}
\includegraphics[width=6.4cm]{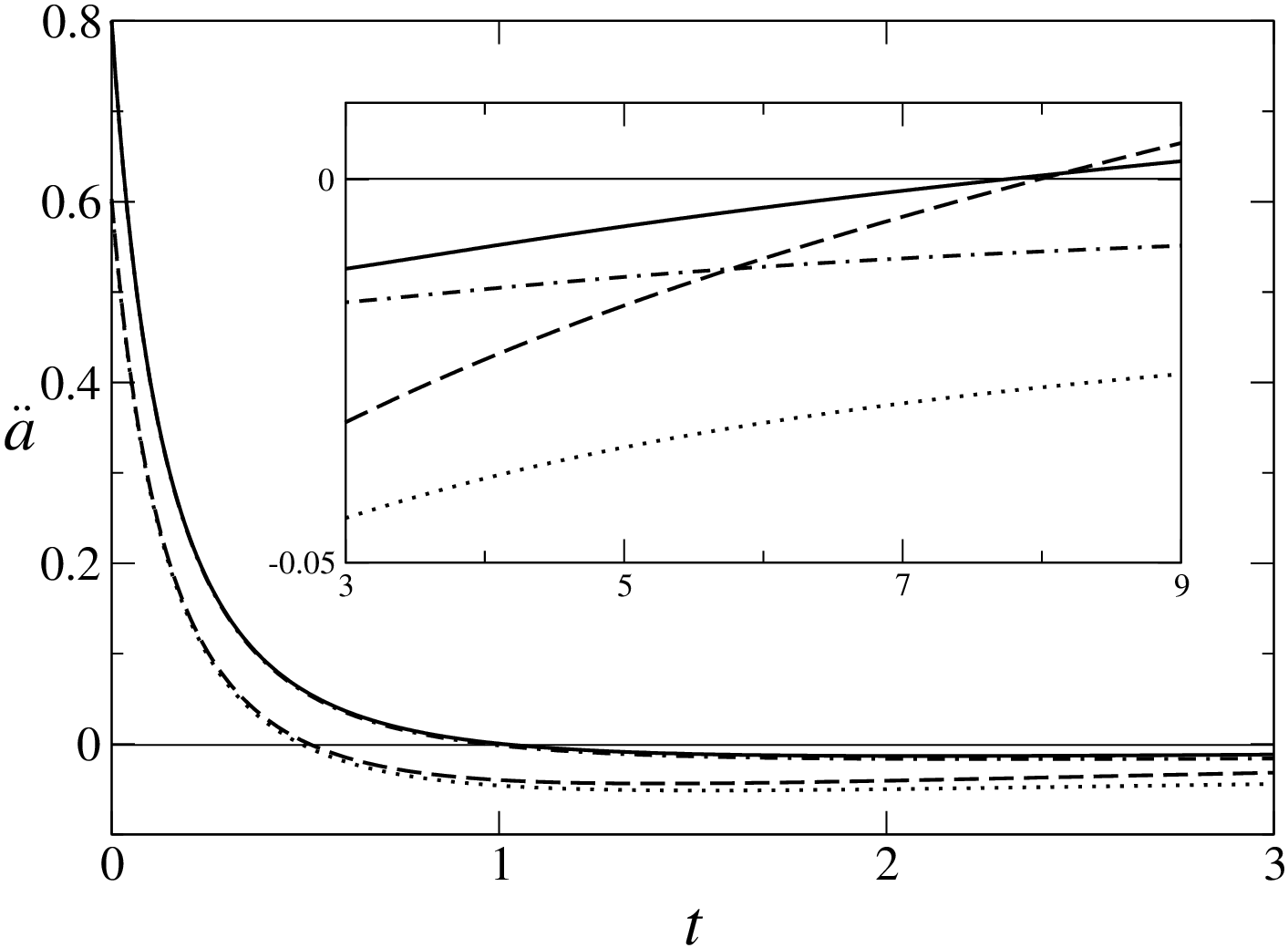}
\caption{Acceleration  vs time  for $\alpha=0.5$ and a) $w=0.9, \lambda=0.002$
 -- straight line; b) $w=0.9, \lambda=0$ -- dash-dotted line;
c)  $w=0.8, \lambda=0.002$ -- dashed line; d)  $w=0.8, \lambda=0$ --
dotted line.}
\end{center}\end{figure}
\begin{figure}\vskip0.8truecm
    \begin{center}
      \includegraphics[width=4cm,height=4.5cm]{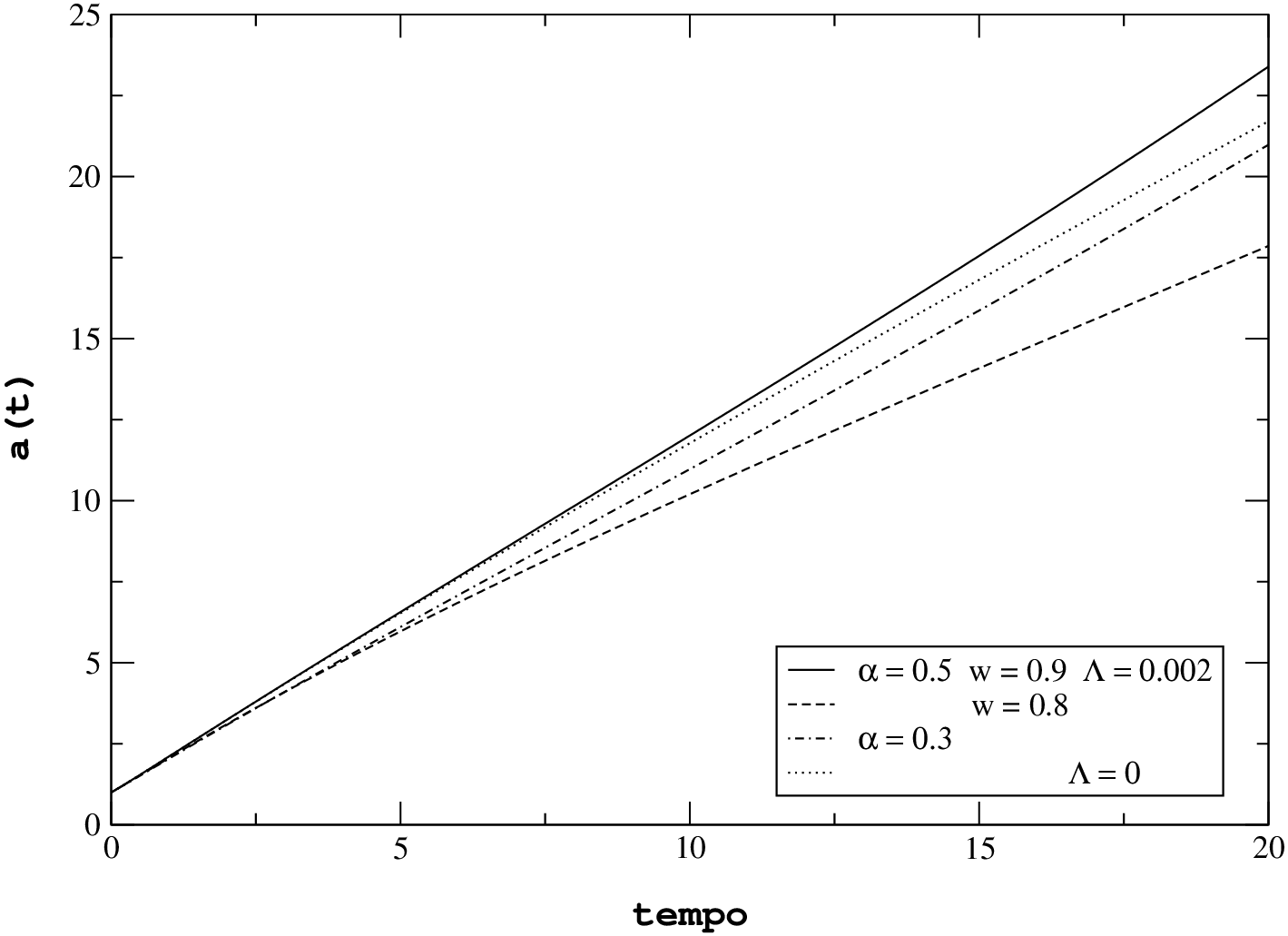}
\hspace*{0.5cm}\includegraphics[width=4cm,height=4.5cm]{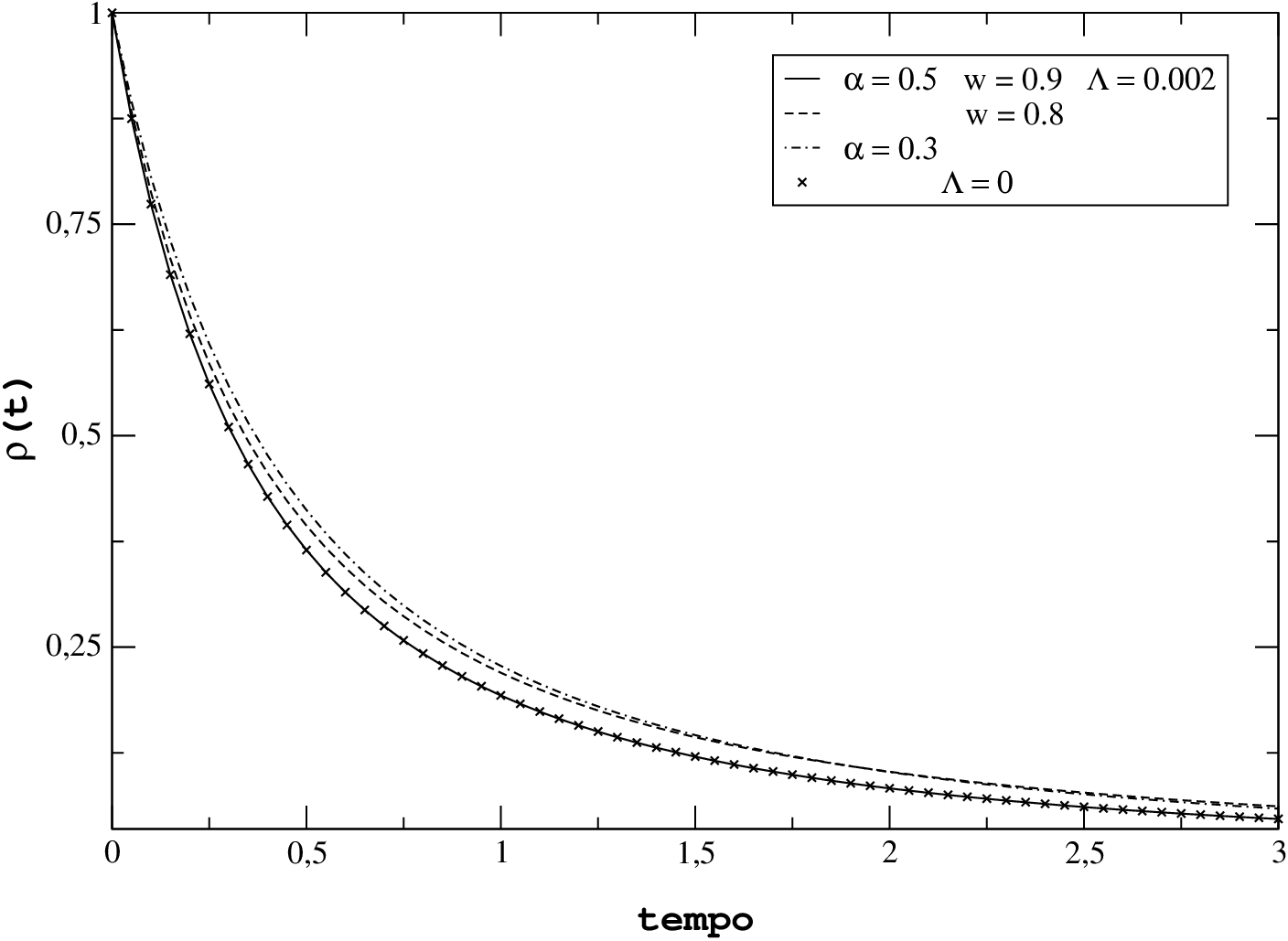}
      \caption{Behavior of cosmic scale factor and energy density
vs time for a) $\alpha=0.5$, $\lambda=0.002$, $w=0.9$ -- 
straight line, b) $\alpha=0.5$, $w=0.8$, $\lambda=0.002$ -- dashed
line; c) $\alpha=0.3$, $w=0.9$, $\lambda=0.002$ -- 
dash-dotted line; d) $\alpha=0.5$, $w=0.9$, $\lambda=0$ --
dotted line for $a(t)$ and crosses for $\rho(t)$.}
    \end{center}
  \end{figure}

We have also investigated the accelerated regimes
 that follow from the
CGHS model by considering  again the  VDW  equation of state.
 We have solved numerically the system of differential 
equations  (\ref{99}), (\ref{102}) and 
(\ref{103})  for the energy density, the cosmic scale factor and the dilaton, 
respectively.
The initial  conditions chosen were: 
$ \rho(0)=1, a(0)=1,\dot a(0)=1, \phi(0)=0, \dot \phi(0)=d  $
were  $d$ is a constant that  determine the transition from
 the different regimes
obtained, as we discuss bellow.
 The behavior of the acceleration depends strongly on
the initial condition for the time derivative of the dilaton $d$. In fact,
the possibility of describing a transition from an accelerated period and
a decelerated one starts to make sense when we take values  of $d<-0.62$.
The time evolution  of the cosmic scale
factor and of its acceleration field show a dramatically
different behavior when compared to the ones obtained in the JT model.
The acceleration vanishes for large values of $t$ even with a non-vanishing 
cosmological constant. Furthermore, a vanishing cosmological
constant implies  that the cosmic scale factor tends to a constant value, 
i.e., the 2D Universe tends to a static Universe. For a non-vanishing
cosmological constant the solution shows a 2D
Universe in permanent  expansion with the energy density decreasing
accordingly. 
Again, as the Universe is expanding the VDW
equation approaches a barotropic equation of state.
 For a fixed value of the parameter $w$  the results show that the
smaller the
value of $\alpha $ the more drastic is the transition.
On the other hand, 
for a fixed $\alpha$ the smaller the value of $w$ the transition to the   
decelerated regime gets more drastic.  In all cases, after the transition
occurs, the accelerated regime never returns.

\newpage

\end{document}